\documentclass[11pt]{optica-article}
\usepackage[utf8]{inputenc}

\journal{opticajournal} % for journals or Optica Open

\articletype{Research Article}

\usepackage{lineno}
\usepackage{algorithm}
\usepackage{algpseudocode}

\usepackage{amsmath}

\usepackage{graphicx}   % figure placement
\usepackage{booktabs}   % nicer tables
\usepackage{textcomp}   % registered and copyright symbol

\DeclareMathOperator*{\argmin}{arg\,min}
\newcommand{\appropto}{\mathrel{\vcenter{
  \offinterlineskip\halign{\hfil$##$\cr
    \propto\cr\noalign{\kern2pt}\sim\cr\noalign{\kern-2pt}}}}}
\begin{document}

\title{Bayesian uncertainty evaluation applied to the tilted-wave interferometer}

\author{Manuel Marschall,\authormark{1,*} 
        Ines Fortmeier,\authormark{2}
        Manuel Stavridis,\authormark{1}
        Finn Hughes,\authormark{1}
        and Clemens Elster\authormark{1}}
\address{\authormark{1}Physikalisch-Technische Bundesanstalt (PTB), Abbestra{\ss}e 2-12, 10587 Berlin, Germany.\\ 
        \authormark{2}Physikalisch-Technische Bundesanstalt (PTB), Bundesallee 100, 38116 Braunschweig, Germany.}
\email{\authormark{*}manuel.marschall@ptb.de}

\begin{abstract*}
    The tilted-wave interferometer is a promising technique for the development of a reference measurement system for the highly accurate form measurement of aspheres and freeform surfaces. The technique combines interferometric measurements, acquired with a special setup, and sophisticated mathematical evaluation procedures. To determine the form of the surface under test, a computational model is required that closely mimics the measurement process of the physical measurement instruments. The parameters of the computational model, comprising the surface under test sought, are then tuned by solving an inverse problem. Due to this embedded structure of the real experiment and computational model and the overall complexity, a thorough uncertainty evaluation is challenging. In this work, a Bayesian approach is proposed to tackle the inverse problem, based on a statistical model derived from the computational model of the tilted-wave interferometer. Such a procedure naturally allows for uncertainty quantification to be made. We present an approximate inference scheme to efficiently sample quantities of the posterior using Monte Carlo sampling involving the statistical model. In particular, the methodology derived is applied to the tilted-wave interferometer to obtain an estimate and corresponding uncertainty of the pixel-by-pixel form of the surface under test for two typical surfaces taking into account a number of key influencing factors. A statistical analysis using experimental design is employed to identify main influencing factors and a subsequent analysis confirms the efficacy of the method derived. 
\end{abstract*}

\section{Introduction}
\label{sec:Introduction}
Aspheres and freeform surfaces enable smaller and lighter optics with even better optical properties and are therefore in high demand in the optics industry~\cite{Braunecker2008}. However, their manufacturing accuracy is limited by the current state of the art of surface form measurement systems and there is a need for traceable optical form measurements with low uncertainty~\cite{Schachtschneider_ea_2018}. Interferometric form measurements belong to the most accurate measurements with high resolutions. But, in contrast to interferometric form measurements of simple forms like flats or spherical surfaces, it is challenging to generate a reference wavefront that matches the ideal surface such that the measured fringe densities directly correspond to the form deviations sought. Therefore, interferometric measurement systems for complex surfaces can be divided into null-test systems, where effort is put into the generation of complex reference wavefronts that correspond to the ideal surface under test (SUT)~\cite{Pruss2004}, stitching methods~\cite{Supranowitz2016}, where moving parts are needed and the null-test conditions are fulfilled locally, or non-null test measurement methods, where sophisticated data evaluation methods are needed to reconstruct the surface form from the data measured. At the Physikalisch-Technische Bundesanstalt (PTB), the national metrology institute of Germany, a reference measurement system based on a tilted-wave interferometer (TWI)~\cite{baer2014fast,fortmeier2016evaluation,baer2014calibration} is under development~\cite{fortmeier2022development}. The TWI belongs to the non-null test measurement methods. The measurement principle of the TWI combines interferometric measurements with ray tracing, perturbation methods and mathematical evaluation procedures.

A key aspect of the measurement principle of the TWI is that a \emph{computational model (CM)} is required to obtain the measurand desired~\cite{fortmeier2022development}, i.e., the form of the SUT. This CM, also referred to as \emph{forward model}, closely resembles the measurement process of the TWI. This includes the optical system (lenses, apertures, camera, etc.), ray-tracing and ray-aiming algorithms, and a model for the topography of the SUT. In the CM of the TWI, this topography is parameterized by an a priori known design topography and a difference topography. During the measurement process, the measurements of the real-world TWI are used to adjust the parameters of the difference topography in the CM by solving an inverse problem. This embedded structure of the TWI and its CM to obtain the measurand challenges the application of classical approaches to measurement uncertainty, e.g., the techniques considered in the ``Guide to the expression of uncertainty'' (JCGM 100)~\cite{JCGMGUM} or its supplements~\cite{JCGM101,JCGM102} (JCGM 101/102).

A well-known statistical approach to inverse problems that is also discussed in the JCGM GUM-6~\cite{jcgm-2020gum6} is the Bayesian paradigm. Based on a statistical model for the observations and prior knowledge about its parameters, a state-of-knowledge posterior distribution in terms of a probability density function (PDF) is obtained that allows for probability statements, conditional on the data observed. To this end, Bayes Theorem is applied to update the prior under new observations, hereby allowing the prior to incorporate expert knowledge about parameters of the statistical model, which usually requires elicitation techniques~\cite{o2019expert}. For the TWI, previous publications have already analyzed and quantified certain parameters using sensitivity analysis~\cite{fortmeier2013sensitivity,fortmeier2014results} and experimental design~\cite{scholz2022experimental}. Also, surrogate (black-box) modelling~\cite{beisswanger2023retrace} and first Monte Carlo simulation studies have been performed~\cite{harsch2019monte,schindler2020methoden} to determine parameter sensitivities and main uncertainty sources~\cite{baer2017diss}. However, a full and realistic uncertainty estimation for the SUT is still a challenging task and subject to current research. This is mainly due to the complexity and high-dimensionality of the problem and the embedded CM. Tracing millions of rays through a virtual optical system is already numerically challenging and performing a subsequent uncertainty evaluation increases the computational costs, e.g., when Markov Chain Monte Carlo~\cite{robert1999monte} needs to be employed to obtain samples from the posterior.

This work applies the Bayesian approach to the inverse problem of the TWI and to the task of uncertainty estimation. To this end, a statistical model and informative prior knowledge about the parameters of the model is derived, both of which were derived from the CM. To tackle the computational complexity, an approximate Bayesian inference for the posterior PDF is developed that allows for efficient independent Monte Carlo sampling to compute the uncertainty corresponding to the unknown parameters. The developed approach is then applied to the TWI, by taking into account a number of key influencing factors, to demonstrate the performance of the methodology. This is, to the best of our knowledge, the first Bayesian approach applied to the TWI to obtain an uncertainty estimate of the form of the SUT. Moreover, the statistical and numerical approach presented in this work is generic in the sense that its application is not limited to the TWI. The algorithmic description may also facilitate other measurement procedures that involve or rely on computational models to mimic a real measurement process.

This paper is organized as follows. Section~\ref{sec:setting} introduces the calibration and measurement principle of the TWI, and highlights the embedded CM as an instrument to obtain an estimate. Subsequently, section~\ref{sec:mathematical model} considers the inverse problem from a statistical point-of-view and applies the Bayesian approach to tackle the inference. Numerical application and evaluation for typical surfaces are performed in section~\ref{sec:numerics}. Conclusions, future research questions and a discussion is given in the final section~\ref{sec:discussion}.

\section{The tilted-wave interferometer measurement principle}
\label{sec:setting}
A detailed description of the measurement setup and its evaluation principle can be found in literature~\cite{baer2014fast,baer2014calibration,fortmeier2016evaluation,fortmeier2022development}, but, for the sake of convenience, it is briefly recalled in the following. The basic measurement principle of the TWI combines a special interferometric measurement setup with model based evaluation procedures. The principle can be divided in three phases: the calibration (i.e., the adjustment of the CM to the experimental setup), the acquisition of measurement data, and the mathematical evaluation of the SUT. In this section, each phase is briefly discussed and the notation required is introduced. Also, important aspects for the subsequent estimation of uncertainty are highlighted.

\subsection{Interferometric measurements and the computational model of the TWI}
 
\begin{figure}
    \centering
    \includegraphics[width=.7\linewidth]{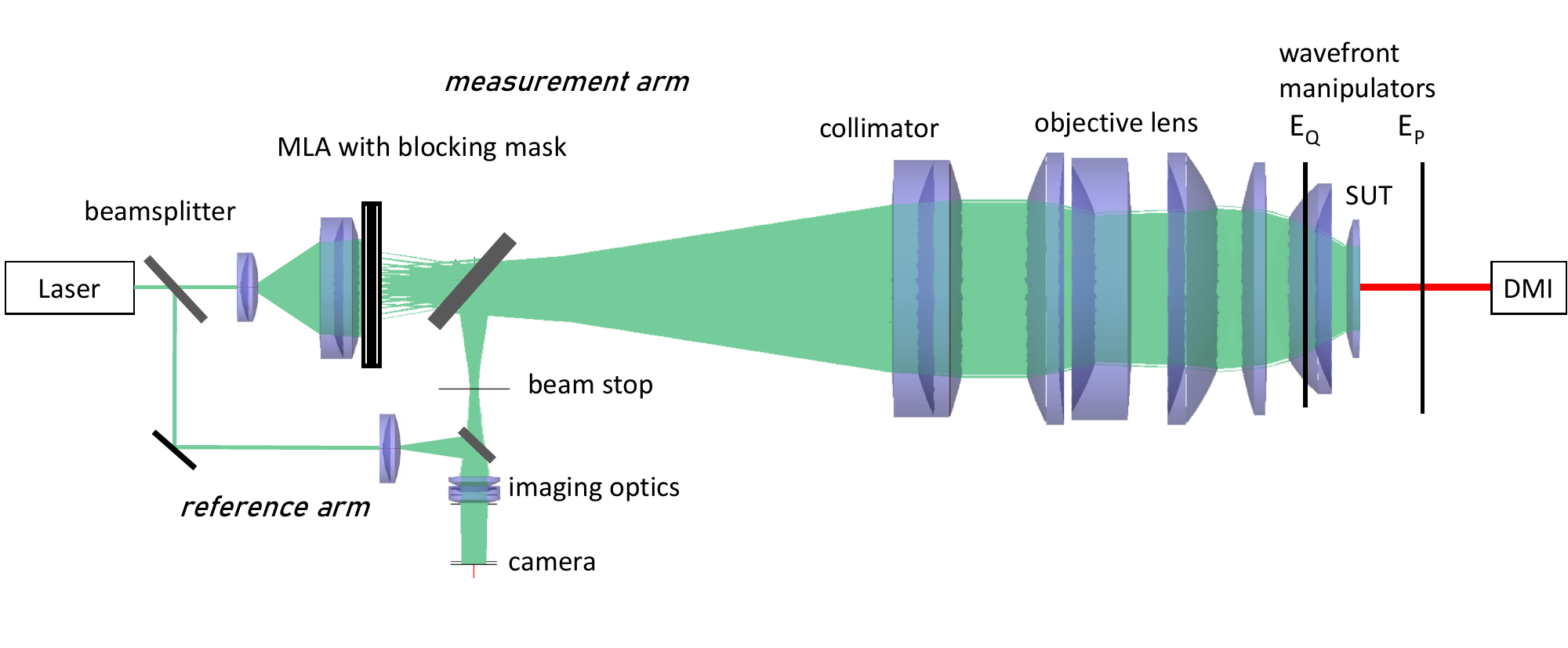}
    \caption{Sketch of the basic setup of the TWI (adjusted from \cite{fortmeier2022development}, CC BY 4.0).}
    \label{fig:setupTWI}
\end{figure}

The basic measurement setup of the TWI is shown in Figure~\ref{fig:setupTWI}. The light of a laser is split by a beamsplitter into the reference arm and the measurement arm of the interferometer. The measurement arm is equipped with a 2D microlens array, where each microlens acts as a single point light source. With this setup, each microlens generates a differently tilted wavefront behind the collimator that illuminates the SUT. A beam stop in the Fourier plane of the imaging optics avoids sub-sampling effects by blocking the light that would produce non-resolvable fringe densities at the camera. Depending on the local slope of the specimen, the light from a different microlens generates resolvable sub-interferograms, referred to as patches, at the camera. These patches contain information about the form of the SUT.
To avoid interference between the light of neighbouring microlenses, a special mask behind the microlens array blocks the light from every second row and column of the 2D microlens array, so that four measurements with four different mask positions are needed to acquire the images from all possible microlenses. For each of the four measurements, a five-step phase shifting algorithm~\cite{Hariharan1987} together with the Goldstein unwrapper~\cite{Goldstein1988} are used to calculate the optical path length differences (OPDs), i.e., the differences between the optical path lengths (OPLs) of the reference arm and the measurement arm, from the interferograms. Since the positioning of the SUT within the measurement setup, especially along the optical axis, is important, the setup is equipped with an additional distance-measuring interferometer, which measures the relative position of the SUT during the alignment procedures. In order to reconstruct the form of the SUT from these OPDs, a CM of the setup and the SUT is used to model the OPDs for the assumed surface form. By solving an inverse problem, the parameters of the assumed surface form are adjusted in an iterative process, until the measured and modeled OPDs match to each other.

For this purpose, the measurement concepts of the TWI are replicated in a virtual environment using the optical simulation toolbox SimOptDevice~\cite{schachtschneider2019simoptdevice}. Mathematically, this defines a numerical function $g(\theta, Z)$ that models OPDs that closely resemble the OPDs from the real-world TWI. We will refer to $g(\theta, Z)$ as the CM of the TWI. Here, the parameter $\theta$ describes an a priori unknown value for the parameterization of the form of the SUT together with its position and orientation within the setup. The parametrization chosen is given by the design description (for an asphere, e.g., the aspherical formula of~\cite{pruss2008testing,ISO.10110-12:2019}) and Zernike polynomials~\cite{wang1980wave}, which are commonly employed to describe variations in optics. Moreover, $\theta$ comprises a set of latent variables that are required for an efficient solution of the following inverse problem. In particular, each patch of the acquired interferograms has an offset value corresponding to a shift of the OPDs of that patch, since the OPDs of each patch can only be calculated from the measurement data up to an unknown offset value. Since the actual parametrization and dimensionality of $\theta$ depends on the form of the SUT, a discussion for two typical SUTs is presented in the corresponding sections of~\ref{sec:numerics}. The multivariate parameter $Z$ collects additional parameters of the experiment that are required to replicate the TWI measurements in a virtual environment, i.e., the configurations of the lens-systems, the position of the collimator and CCD, the wavelength of the laser employed, to name just a few. For details on the elements of $\theta$ and $Z$, cf. e.g.~\cite{fortmeier2022development}. 

\subsection{Evaluation of the estimate for the measurand}
\label{sec:measurement}
The CM is required to estimate the form of the SUT. In particular, since the TWI is an indirect measurement device, the multivariate parameter $\theta$ of the CM is adjusted, such that the resulting modeled OPDs closely resemble the OPDs $y$ acquired by the real TWI. Mathematically, this process is formulated as an inverse problem. Due to the complex structure of the real experiment and hence also of the CM, this problem is non-linear and numerical optimization is required to obtain a suitable solution. Moreover, the number of rays selected to solve the inverse problem is much larger than the number of parameters contained in $\theta$, which may render the inverse problem also ill-conditioned or even ill-posed.

A common approach considered to tackle inverse problems is non-linear least-squares estimation~\cite{hartley1965nonlinear}. For our purpose, the following loss function is minimized for some fixed parameter $\hat Z$ that describes the current estimate of the model parameters, e.g., position and orientation of lenses, the position of the collimator and CCD, the laser wavelength, etc.,
\begin{equation}
  \label{eq:leastsquaresestimate}
    \hat \theta_{\hat Z} \in \argmin_{\theta} \|y - g(\theta, \hat Z)\|_2^2.
\end{equation}
In general, there is no analytical solution available to~\eqref{eq:leastsquaresestimate}. To obtain a numerical solution to the inverse problem, repeated evaluation of the CM for different values of $\theta$ is required until an ``optimal'' value is obtained. In this work, the parametrization of $\theta$ slightly deviates from the one in~\cite{fortmeier2022development}. In particular, orientation parameters are removed from the estimation to ensure that the Jacobian of the least-squares functional~\eqref{eq:leastsquaresestimate} is of full rank. A detailed discussion is given in appendix section~\ref{sec:appendix:parameter}.

As previously mentioned, the parametrization that describes the deviation of the SUT from its assumed form in $\theta$ is given by Zernike polynomials. Since these polynomials are smooth, they are not able to reflect mid- and high spatial frequencies of typical manufacturing errors of the surface. Therefore, the parameter $\theta$ is an intermediate result and the measurement principle of the TWI requires a subsequent correction of positioning and orientation of the parametric description of the SUT. To this end, a subsequent data analysis stage is applied which transforms the parametric description of the SUT into a non-parametric, pixel-wise representation, cf. e.g.~\cite{baer2013measurement} and~\cite[section 3.3.2]{fortmeier2022development}.
This post-processing step is formally denoted by the function $\mu = f(\theta, Z)$ that, given values for $\theta$ and $Z$, estimates a 3D point-cloud $\mu$, which describes the SUT pixel-wise. In metrology, this quantity $\mu$ is usually referred to as the measurand. Since the post-processing step corrects positioning and orientation according to an assumed design topography, the uncertainty for the measurand is expected to be much smaller than the uncertainty of the polynomial coefficients $\theta$.

To tackle uncertainty evaluation for this two-stage procedure, a corresponding two-step approach will be developed in section~\ref{sec:bayesian inference}. For the non-linear regression problems of the form~\eqref{eq:leastsquaresestimate}, many statistical approaches exist to estimate uncertainties. An overview and possible approximation techniques in the context of measurement uncertainty are given, e.g., in~\cite{malengo2013weighted}. In section~\ref{sec:mathematical model}, a statistical modeling approach using the Bayesian paradigm is presented. For the final transformation of the auxiliary, parametric representation of the SUT $\theta$ into the actual measurand $\mu$, a propagation of PDFs approach as considered in the JCGM-101~\cite{JCGM101} is performed. 

\subsection{Calibration of the computational model of the TWI}
\label{sec:calibrationVE}
An important step required for highly accurate estimation of the measurand is the calibration, i.e., the adaptation of the model of the interferometer used in the CM to reflect the behavior of the real experiment as closely as possible. During the calibration process, the CM of the TWI is fine-tuned by a phenomenological correction of the OPDs with the help of two wavefront manipulators~\cite{fortmeier2022development}, cf. Figure~\ref{fig:setupTWI}. These wavefront manipulators are parametrized by a product of Zernike polynomials and the coefficients are estimated, such that the CM of the TWI generates OPDs that closely resemble actually measured OPDs from two well-known reference spheres measured at a large number of positions in the test space. The large number of positions of the reference spheres is chosen with the intention to correct all the beams passing through the interferometer. 

Mathematically the calibration, similar to the evaluation of the measurand, is an inverse problem. However, the role of the parameters in the CM are exchanged, since the parametrization of the reference spheres is well-known and the calibration parameters of the CM are to be determined. For more details on the calibration itself we refer to~\cite{fortmeier2022development} and for a discussion on an uncertainty evaluation for the TWI calibration, cf. appendix section~\ref{sec:appendix:calibration}.

\section{Bayesian uncertainty evaluation}
\label{sec:mathematical model}
So far, section~\ref{sec:setting} introduced the measurement principle of the TWI and the procedures required to obtain an estimate for the form of the SUT as a solution of an inverse problem. In this section, uncertainty evaluation is considered to assign a corresponding uncertainty to the estimated value. To this end, a statistical model for the measurements that is based on the CM employed in the TWI is introduced. Then, Bayes Theorem is applied to update some prior knowledge according to the OPDs calculated from the interferograms measured. Subsequently, an approximation to the posterior covariance is introduced. Since this section is generic, it may be applied to other measurement settings, e.g., applications such as scatterometry~\cite{heidenreich2018bayesian} and flow meter measurements~\cite{straka2022simulation} may capitalize from the methods described in the following.

\subsection{Statistical modeling and approximate Bayesian inference}
\label{sec:bayesian inference}
The OPDs $y\in\mathbb{R}^n$ that are acquired by the TWI are statistically modeled as realizations of the following homoscedastic model
\begin{equation}
  \label{eq:statistical model}
    Y\;|\;\theta, Z, \sigma^2\sim N(g(\theta, Z), \sigma^2 I), \quad\sigma^2>0 \;\text{known}.
\end{equation}
The function $g(\theta, Z)$ denotes the CM of the TWI, which mimics the measurement process as a replacement for the actual TWI, $Z$ models additional parameters (e.g., position, orientation and optical properties of the optical elements, wavelength of the laser, etc.) and $\theta$ is the sought parameter (i.e., a parametric description of the form of the SUT in terms of Zernike polynomials). A detailed compilation of involved parameters for the TWI can be found in~\cite{scholz2022experimental, Fortmeier2017StepsTraceability, fortmeier2022development} and appendix section~\ref{sec:appendix:parameter}-\ref{sec:appendix:calibration}. The Gaussian measurement noise variance $\sigma^2$ is assumed to be known. This simplifies computations and the extension to a heteroscedastic model is straight forward. 

The goal is to perform uncertainty quantification for the usually obtained least-squares estimate~\eqref{eq:leastsquaresestimate}, which will be denoted as $\hat\theta_{\hat Z}$ to indicate the use of the estimate $\hat Z$, partially obtained by calibration and expert knowledge. This uncertainty quantification can be accomplished by a Bayesian approach. To this end, consider the prior knowledge $\pi(\theta, Z) = \pi(\theta)\pi(Z)$, which assumes independence between the inference parameter $\theta$ and the additional parameter in the CM\footnote{For $\pi(Z)$ we assume perfect prior knowledge for the parameters that correspond to the optical systems and a prior PDF for the Zernike coefficients of the wavefront manipulators according to appendix section~\ref{sec:appendix:calibration}.} $Z$. Then, using the statistical model~\eqref{eq:statistical model}, the joint posterior PDF is given by
\begin{equation}
  \label{eq:joint posterior}
    \pi(\theta, Z\;\vert\; y) = h(y) \exp\left(-\|y - g(\theta, Z)\|_2^2/2\sigma^2\right)\pi(\theta)\pi(Z),
\end{equation}
where the constant $h(y)$, i.e., the evidence, normalizes the posterior PDF, such that it integrates to one. Throughout this work, it is assumed that this PDF exists. For more details and requirements on the function $g$ to ensure existence, propriety and the existence of moments for the posterior, cf. eg.~\cite{gelman1995bayesian,stuart2010inverse}. Marginalization over the parameter $Z$ yields the marginal posterior PDF for the parameter of interest $\theta$. This PDF reflects the state-of-knowledge about this parameter after consideration of the data $y$. Now, the covariance of the estimate $\hat\theta_{\hat Z}$ with respect to this marginal posterior is the quantity that describes the sought uncertainty under the available state-of-knowledge. This covariance matrix is given by
\begin{align}
  \label{eq:covariance}
  U &= \mathbb{E}_{\theta\vert y}\left[(\theta - \hat\theta_{\hat Z})(\theta - \hat\theta_{\hat Z})^T\right] \\ 
  &= h(y)\int\int (\theta - \hat\theta_{\hat Z})(\theta - \hat\theta_{\hat Z})^T\exp\left(-\|y - g(\theta, Z)\|_2^2/2\sigma^2\right)\mathrm{d}\pi(\theta)\mathrm{d}\pi(Z), \nonumber
\end{align}
which requires the computation of a multivariate integral over a non-linear function involving the CM, which is in general not feasible and a suitable approximation is required. Therefore, a partial first order Taylor expansion of the CM in the parameter $\theta$ is performed: $g(\theta, Z)\approx g(\hat\theta, Z) + J(\hat\theta, Z)(\theta - \hat\theta)$, where $J(\hat\theta, Z)$ denotes the Jacobian matrix of the function $g$ around some expansion point. This linearization is considered for all possible values of $Z$ and the expansion point is given by
\begin{equation}
  \label{eq:hat theta}
    \hat\theta = \hat\theta(Z) \in \argmin_{\theta} \|y - g(\theta, Z)\|^2_2.
\end{equation}
It is to note that $\hat\theta$ does not usually equal $\hat\theta_{\hat Z}$. The linearization allows analytical integration with respect to $\theta$ in the integral~\eqref{eq:covariance}, assuming a constant (non-informative) prior for the measurand $\pi(\theta)\propto 1$ and a full-rank Jacobian matrix for every $Z$. Then, the covariance matrix desired can reasonably be approximated by
\begin{equation}
    \label{eq:covariance monte carlo}
    U\approx \int \sigma^2\left(J(\hat\theta, Z)^TJ(\hat\theta, Z)\right)^{-1} + (\hat\theta - \hat\theta_{\hat Z})(\hat\theta - \hat\theta_{\hat Z})^T\mathrm{d}\pi(Z),
\end{equation}
where the quality of the approximation depends on the quality of the Taylor approximation. The covariance matrix $U$ can therefore be obtained by Monte Carlo sampling. A detailed derivation of the formula~\eqref{eq:covariance monte carlo} is given in appendix~\ref{sec:details covariance}. It should be noted that the choice of a non-informative prior for $\theta$ is for mathematical convenience and can be replaced by a vague prior with large variance without significantly affecting the resulting uncertainty.

To numerically stabilize the computations and reduce the complexity of the Jacobian matrix $J(\hat\theta, Z)$ that has to be assembled for every $Z$, a decomposition is performed. To accelerate the computations in~\eqref{eq:covariance monte carlo}, a singular value decomposition of the Jacobian is chosen to be $\tilde{U} \Lambda V^T = J(\hat\theta, Z)$, since the required matrix inversion simplifies to
\begin{equation}
    \left(J(\hat\theta, Z)^TJ(\hat\theta, Z)\right)^{-1} = V \Lambda^{-2} V^T.
\end{equation}
Hence, computations can be limited to the orthogonal matrix $V$ and the singular values of the Jacobian, which are, by construction of the measurement procedure outlined in section~\ref{sec:measurement} and appendix section~\ref{sec:appendix:parameter}, always greater than 0.

The uncertainty estimation procedure for the TWI derived above is outlined in Algorithm~\ref{alg:mc}. Note again that this algorithm can be employed for similar applications, involving models as~\eqref{eq:statistical model}, just as well.

\begin{algorithm}[H]
\caption{Approximate Bayesian uncertainty evaluation}\label{alg:mc}
\begin{algorithmic}[1]
\Require  Computation model $g(\theta, Z)$, prior $\pi(\theta, Z) \propto \pi(Z)$, number of samples $N$.
\Ensure Approximate covariance for estimate $\hat\theta$ according to~\eqref{eq:covariance monte carlo}.
\State Perform a calibration to estimate $\hat Z$ \Comment{Usually expensive but required only once, cf. section~\ref{sec:calibrationVE} and~\ref{sec:appendix:calibration}}
\State Perform a measurement to estimate $\hat\theta_{\hat Z}$ \Comment{Usually performed to obtain an estimate, cf. section~\ref{sec:measurement}.}
\For{$i = 1, \ldots, N$}
  \State $Z_i \leftarrow$ Draw from prior $\pi(Z)$ 
  \State $\hat\theta_i \leftarrow$ Perform a measurement to obtain current estimate using a disturbed CM~\eqref{eq:hat theta}.
  \State $J(\hat\theta_i, Z_i)\leftarrow$ Calculate Jacobian at current estimate $(\hat\theta_i, Z_i)$.
  \State $U_i \Lambda_i V_i^T = J(\hat\theta_i, Z_i)\leftarrow$ Perform singular value decomposition.
  \State Store matrix $V_i$ and vectors $\hat\theta_i$ and $\Lambda_i$ in memory.
\EndFor
\State Compute approximate covariance matrix
\begin{equation}
    U \approx \frac{1}{N} \sum_{i=1}^N \sigma^{2} V_i\Lambda_i^{-2} V_i^T + (\hat\theta_i - \hat\theta_{\hat Z})(\hat\theta_i - \hat\theta_{\hat Z})^T
\end{equation}
\end{algorithmic}
\end{algorithm}

\subsection{Details for the TWI and its computational model}
\label{sec:DetailsForTWI} %IF label fehlte
For the TWI, the Jacobian matrix required in~\eqref{eq:covariance monte carlo} is available analytically~\cite{fortmeier2014analytical} and automatically obtained by the CM. To this end, the chain-rule is employed similarly to the concept of backpropagation in machine learning~\cite{hecht1992theory}. However, this matrix is large and usually badly-conditioned. In the current realization considered for this work, the number of elements of the Jacobian matrix during calibration is approximately $5\,000\times 250\,000$ with 75\% being non-zero entries. Storing this matrix requires roughly $8.0$ GB of memory and the matrix has a condition number of $10^5$. These figures depend on the resolution of the CM, i.e., the number of rays used for solving the inverse calibration problem (and traced through the modeled optical system). 

For the lateral resolution of the pixel-wise uncertainty, a smaller grid of $350\times 350$ has been chosen, instead of the original non-equidistant resolution which results from 4 camera images with a resolution of $2\,048\times 2\,048$ pixels projected to the specimen, to reduce computational complexity. However, this reduction has negligible impact on the uncertainty. Moreover, we like to mention that the linearization performed to obtain~\eqref{eq:covariance monte carlo} is reasonable for the TWI, since the OPDs acquired and considered by the TWI can be expected to change linearly for deviations of the SUT in the region of a well-chosen estimate.

\subsection{Post-processing and uncertainty estimation for the measurand}
\label{sec:postprocessing}
As described in section~\ref{sec:measurement}, the estimated parameter vector $\hat\theta_{\hat Z}$ is only an auxiliary quantity. In optical surface metrology, one is usually interested in a 3D point cloud representation of the SUT. To this end, the already mentioned post-processing $\mu = f(\theta, Z)$ is employed. Here, $\mu$ denotes the actual measurand, i.e., a pixel-wise representation of the SUT and one is interested in an estimate $\hat\mu$ and its corresponding uncertainty or covariance. The simplification described in appendix section~\ref{sec:appendix:parameter}, in particular not taking the projection of the residuals to consider high-frequency terms into account, renders the measurement model function $f$ independent of the variable $Z$, i.e., $\mu = f(\theta)$. Hence, to obtain an estimate and corresponding uncertainty for the measurand $\mu$, one can safely apply the Monte Carlo procedure for multivariate measurands given by JCGM 102~\cite{JCGM102}. In particular, realizations $\theta_i$ are drawn from the distribution assigned to $\theta$ and subsequently the post-processing $\mu_i=f(\theta_i)$ generates samples from the distribution of the measurand. For the state-of-knowledge PDF of the input quantity $\theta$, we assume that $\theta$ follows a normal distribution with its mean given by the estimate of~\eqref{eq:leastsquaresestimate}, i.e., $\hat\theta_{\hat Z}$ and covariance $U$ from~\eqref{eq:covariance} as computed by Algorithm 1. 

For the general case of $\mu=f(\theta, Z)$, sampling from the joint posterior~\eqref{eq:joint posterior} is required, which is usually more complex. Suitable sampling strategies may require Markov Chain Monte Carlo procedures, which are not considered here. Alternatively, further linearization and approximation steps could be employed to additionally obtain a posterior PDF for the parameter $Z$.

\section{Practical uncertainty evaluation for two typical surfaces}
\label{sec:numerics}
In this work, we present a proof of concepts study for the Bayesian uncertainty evaluation framework presented and study its efficiency for the TWI. To this end, we consider in this section two practical relevant surfaces: an asphere and a toroid which were also analyzed in~\cite{fortmeier2022development}. For each SUT, a calibration of the TWI is performed using two reference spheres. In the current setup at PTB, the first sphere has a radius of  $R_{40} = $ 40.00443 mm, $u_{R_{40}} (k = 2) = $ 0.00020 mm and is measured in 123 positions and the second sphere has a radius of  $R_{15} = $ 14.99531 mm, $u_{R_{15}} (k = 2) = $ 0.00023 mm and is measured in 16 positions of the test space. The positions chosen for calibration also depend on the measurement position of the SUT and are chosen such that all relevant optical paths through the TWI are covered. Hence, for the asphere and for the toroid, different calibrations are used. Both reference spheres have sphericities smaller than the specified value of $\frac{\lambda}{50} \leq 15$ nm root-mean-squared (rms). Note that the uncertainties of the radii and sphericities of the spheres are neglected in this work. The calibration procedure of section~\ref{sec:calibrationVE} and appendix section~\ref{sec:appendix:calibration} generates the estimate for the phenomenological wavefront manipulators, comprising a set of $66\times 28 + 28\times 66 = 3696$ Zernike coefficients. In addition and according to the procedure outlined in appendix section~\ref{sec:appendix:calibration}, a multivariate Gaussian PDF is obtained for the parameter of the wavefront manipulators. The calibrations considered for both specimens are taken from real measurements for which we assume a Gaussian measurement noise on the OPDs with standard deviation of 10 nm.

For the subsequent measurement of the SUT, the complete list of input parameters under consideration in this work, i.e. to demonstrate the method for uncertainty estimation developed, and their considered uncertainty are given in Table~\ref{tab:parameter}. For the calibration parameters it is to note that the multivariate Gaussian distribution comprises the complete set of Zernike coefficients for the wavefront manipulators, i.e., the distribution is $3696$ dimensional. The uncertainty estimation, i.e. the computation of the covariance matrix~\eqref{eq:covariance monte carlo}, is given by the procedure in Algorithm~\ref{alg:mc} for which we employ $N=1000$ samples. In particular, the estimation procedure involving the CM of the TWI has been applied $1000$ times, which takes about two days on a workstation with 256 CPU cores, 1 TB of memory and parallelization across 20 processes using Matlab\textsuperscript{\textregistered}-2023. The Monte Carlo propagation in the post-processing using 2000 independent samples takes roughly 20 min. The dimensionality of the covariance matrix $U$ depends on the underlying SUT. In general, Zernike polynomials in 2D of order up to 20 are considered, ignoring offset and tilts, i.e., up to 228 coefficients are to determine. In addition, a number of patch offset values are required, depending on the SUT. Details are given in the corresponding sections.

\begin{table}[ht]
  \centering
  \caption{List of parameters that are subject to uncertainties and that are considered in this work. For the Gaussian distributions, the mean and standard uncertainty, i.e., standard deviation are shown. For the multivariate Gaussian distribution assigned to the parameters of the wavefront manipulators, the mean value $Z_\mathrm{wav}$ and the covariance matrix $\Gamma_{\mathrm{wav}}$ are derived in appendix section~\ref{sec:appendix:calibration}.\\[1ex]}
  
  \begin{tabular}{lccrr}
    Description of parameter & Symbol & Distribution & mean & std. uncertainty \\
    \toprule
    Deviation of SUT position in x-direction & $\Delta x$ & Gaussian & $0$ m & $5\times10^{-6}$ m \\
    Deviation of SUT position in y-direction & $\Delta y$ & Gaussian & $0$ m & $5\times10^{-6}$ m \\
    Deviation of SUT position in z-direction & $\Delta z$ & Gaussian & $0$ m & $10^{-6}$ m \\
    Deviation of SUT orientation in $\alpha$-direction & $\Delta \alpha$ & Gaussian & $0$ m & $3\times10^{-4}$ rad \\
    Deviation of SUT orientation in $\beta$-direction & $\Delta \beta$ & Gaussian & $0$ m & $3\times10^{-4}$ rad \\
    Deviation of SUT orientation in $\gamma$-direction & $\Delta \gamma$ & Gaussian & $0$ m & $3\times10^{-4}$ rad \\
    Calibration parameters of wavefront manipulators & $Z_{\mathrm{wav}}$ & multiv. Gaussian & $\hat Z_{\mathrm{wav}}$ & $\Gamma_{\mathrm{wav}}$\\
    \bottomrule
  \end{tabular}
  \label{tab:parameter}
\end{table}

At this point it is to note again that the list of parameters in Table~\ref{tab:parameter} is not exhaustive and many more parameters are known to influence the uncertainty of the TWI, e.g., the stability and wavelength of the laser and the radii of the calibration spheres. Also, a realistic determination of the standard deviation of the influencing factors is challenging and subject to future research. Here, only rough orders of magnitude are chosen that are nevertheless realistic. 

Also note that, while the following specimens have been assessed in real measurements, the measurement data considered here are simulated by the software SimOptDevice based on the real measurement results. In particular, high-frequency deviations of the SUT have been removed and Gaussian measurement noise with standard deviation $\sigma=10$ nm has been added to the OPDs.

\subsection{Aspherical surface}
\label{sec:numerics:asphere}
The steep asphere considered here is made of glass, has a clear aperture of 28 mm, and was also analyzed in an interlaboratory comparison in~\cite[asphere 4]{Schachtschneider_ea_2018}. Details about the mathematical description of the asphere, as well as of the calibration and measurement of this asphere using the TWI at PTB are given in~\cite[section 4.2]{fortmeier2022development}.  Similar to the results presented in the reference, we evaluate the form of the asphere on a surface diameter of 24.7 mm. For the reconstruction parameter $\theta$ of the aspherical surface form, 217 offset parameter and 228 polynomial coefficients are considered.

\begin{figure}[htb]
    \centering
    \includegraphics[width=.99\linewidth]{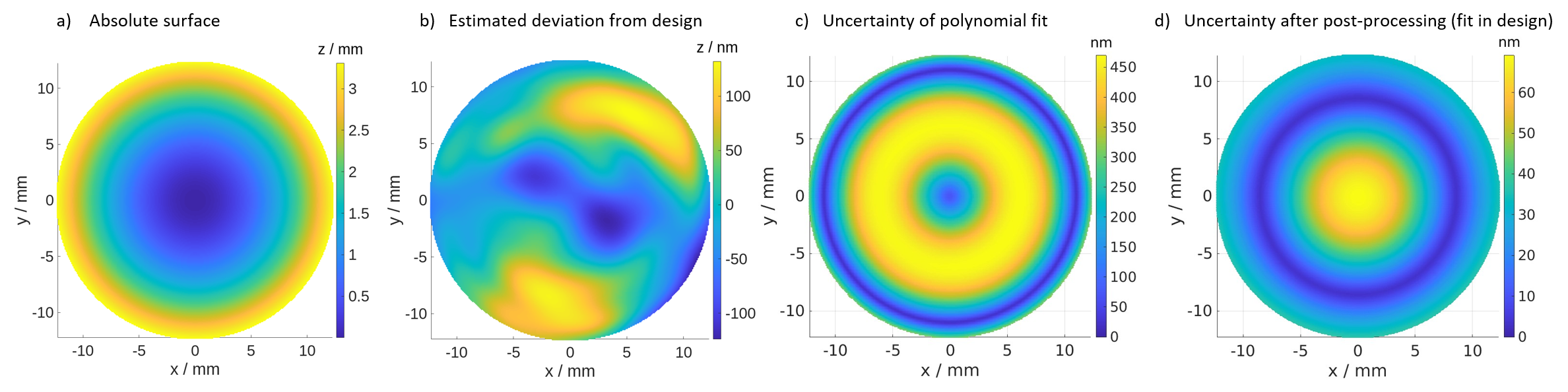}
    \caption{Result for the asphere considered in section~\ref{sec:numerics:asphere}. It is shown in a) the absolute form of the SUT, in b) the estimate of the difference from the design using the original reconstruction procedure from~\eqref{eq:leastsquaresestimate}, in c) the pixel-wise standard uncertainty from the parametric/polynomial fit obtained by the Bayesian approach of section~\ref{sec:bayesian inference} (note that this uncertainty is an intermediate result that does not reflect the actual uncertainty of the TWI estimate), and in d) the result for the pixel-wise standard uncertainty, for the influencing parameters considered, after applying the post-processing step of section~\ref{sec:postprocessing}.}
    \label{fig:asphere}
\end{figure}
In Figure~\ref{fig:asphere} a) we show the estimated absolute form of the surface of the asphere considered and in b) its deviation from the assumed aspherical design, cf.~\cite[section 4.2]{fortmeier2022development}. The intermediate result of the Bayesian uncertainty evaluation procedure derived in section~\ref{sec:bayesian inference} and presented in Algorithm~\ref{alg:mc} is shown in c) of Figure~\ref{fig:asphere}. Hereby, the standard uncertainty shown is obtained by transforming the covariance matrix of the Zernike coefficients obtained by the algorithm to a pixel-wise representation using the linear transformation defined by the Zernike polynomials evaluated at roughly 50000 equidistant points in the circle shown. Subsequently, the square root of the diagonal entries of the resulting matrix is the pixel-wise standard deviation or standard uncertainty of the polynomial fit. Since the positioning and orientation of the SUT is not corrected at this stage, the resulting uncertainty is still large. Note that this result has no practical relevance, but is shown here to demonstrate the two-step approach to generate the final uncertainty estimation of the final measurand. Finally, in d) the standard uncertainty of the measurand, i.e., the asperical surface, obtained after the post-processing step of section~\ref{sec:postprocessing} is shown.

It can be seen in Figure~\ref{fig:asphere} c) that the standard uncertainty of the Zernike fit with up to 450 nm and a rms value taken over all pixels of 320 nm is quite large. As mentioned above, this result is of no practical relevance, since the positioning $\Delta x, \Delta y$ and orientation $\Delta\alpha, \Delta\beta$ is corrected in the post-processing step.  The final resulting standard uncertainty after fitting these parameters is shown in Figure~\ref{fig:asphere} d). Here, the standard uncertainty is reduced to a maximum of 65 nm in the center region and a rms value of 31 nm. The spatial distribution of the final standard uncertainty for the absolute surface form in d) is explained by the remaining dominant uncertainty source of the positioning of the SUT $\Delta z$, which causes a spherical form deviation and has already been identified as a main source of uncertainty in previous publications, cf. e.g.~\cite{fortmeier2016evaluation,scholz2022experimental}. 

Note that for the asphere considered, the parameter $\Delta\gamma$ is considered without uncertainty, i.e. it is fixed to the value 0 rad due to the symmetry of the specimen. Note further that the uncertainty shown in Figure~\ref{fig:asphere} d) is already the pixel-wise standard uncertainty of the absolute surface shown in a), since adding a fixed design as a constant to an uncertainty does not change the result.

\subsection{Convex toroidal surface}
\label{sec:numerics:toroid}
The toroid considered in this section is made of Super Invar\textsuperscript{\textregistered} and has been analyzed for the round-robin comparison in~\cite{fortmeier2020round}. Referring to~\cite{fortmeier2022development}: the specimen has a clear aperture of 50 mm and four Gaussian peak markers of different depths of 0.5, 0.75, 1.0 and 1.25 $\mu$m. Similar to the results presented in the reference, we evaluate the form of the toroid on a diameter of 43.6 mm. For the reconstruction parameter $\theta$ of the toroidal surface form, 21 offset parameter and 228 polynomial coefficients are considered.

\begin{figure}[htb]
    \centering
    \includegraphics[width=.99\linewidth]{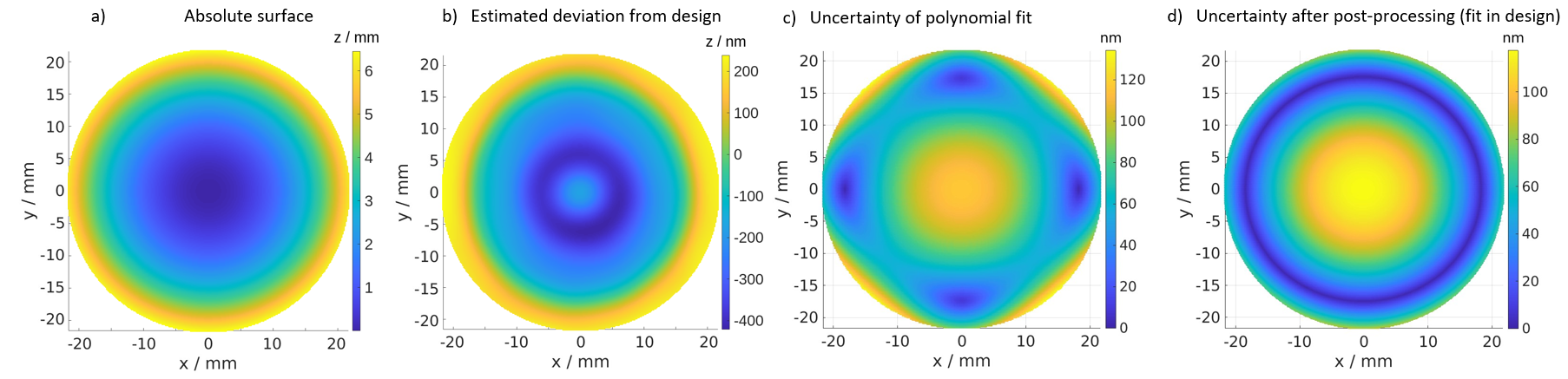}
    \caption{Result for the toroid considered in section~\ref{sec:numerics:toroid}. It is shown in a) the absolute form of the SUT, in b) the estimate of the difference from the design using the original reconstruction procedure from~\eqref{eq:leastsquaresestimate}, in c) the pixel-wise standard uncertainty from the parametric/polynomial fit using the Bayesian approach of section~\ref{sec:bayesian inference} (note that this uncertainty is an intermediate result that does not reflect the actual uncertainty of the TWI estimate), and in d) the result for the pixel-wise standard uncertainty, for the influencing parameters considered, after applying the post-processing step.}
    \label{fig:toroid}
\end{figure}

As in the previous section, in Figure~\ref{fig:toroid} a) we show the estimated absolute surface form of the convex toroidal surface considered and in b) its deviation from the assumed design, cf.~\cite[section 4.1]{fortmeier2022development}. The result of the Bayesian uncertainty evaluation procedure derived in section~\ref{sec:bayesian inference} and presented in Algorithm~\ref{alg:mc} is shown in c) of Figure~\ref{fig:asphere}. Finally, in d) the standard uncertainty of the measurand, i.e., the toroidal surface, obtained after the post-processing step of section~\ref{sec:postprocessing} is shown.

Figure~\ref{fig:toroid} c) shows a large uncertainty in the center and boundary region of the surface with standard uncertainty up to 130 nm and a rms value of 75 nm. As already mentionend for the asphere, this standard uncertainty has no practical relevance and is only shown here for demonstrating the full results of the two-step approach. The subsequent post-processing then reduces the rms value of the standard uncertainty to 63 nm and the final spatial distribution of the standard uncertainty in d) may again be attributed to the important uncertainty source of the surface positioning in $z$-direction.

\subsection{Sensitivity analysis and experimental design}
\label{sec:numerics:sensitivity}
The purpose of experimental design is to determine how influential each of the parameters of an experiment are on the output produced by the model. This can be done through calculating the main effect of individual parameters on the response variable~\cite{saltelli2008global}. In contrast to the work in~\cite{scholz2022experimental}, where the rms value of the form deviation of the reconstructed topography estimate has been analyzed, we consider in this work the rms value of the pixel-wise standard uncertainty as the output. Therefore, for the results in this work, the experimental design analysis is used to underpin our conjectures about the main influencing factors to the uncertainty shown in Figure~\ref{fig:asphere} and Figure~\ref{fig:toroid}. For the following analysis, we consider only the case of the toroid of section~\ref{sec:numerics:toroid} and perform the uncertainty evaluation with a reduced number of Monte Carlo samples $N=200$.

Through experimental design, we are able to identify influential parameters by conducting a series of runs of the experiment where the parameters of the experiment are adjusted to assigned higher or lower levels. For an efficient and thorough analysis, the parameters from Table~\ref{tab:parameter} are extended and summarized in the following groups: $X_1=Z_\mathrm{wav}, X_2=\Delta z, X_3=\mathrm{SUT_{\Delta TOPO}}, X_4=(\Delta x, \Delta y), X_5=(\Delta\alpha,\Delta\beta), X_6=\Delta\gamma$. In particular, we added the parameter $X_3$, which denotes the deviation of the SUT from the expected design, allowing us to analyze the effect of larger design deviations. Moreover, we combined the position parameter in $X_4$ and orientation parameter in $X_5$, since their effect is assumed to be similar. This is, however, not expected for the orientation parameter $\Delta \gamma$, hence we consider $X_6$ individually. In a next step, each parameter $X_1$ to $X_6$ is assigned a number of levels, i.e., different values that can be assigned to it. Since we aim to analyze the impact of the parameter uncertainty to the resulting standard uncertainty of the SUT, we consider different levels of standard uncertainty that is assumed for the parameter of the experiment. In particular, for the calibration parameters of the wavefront manipulators $X_1$, two levels are chosen: either 0 equating to no uncertainty, or 1, which corresponds to the covariance $\Gamma_{\mathrm{wav}}$ given in Table~\ref{tab:parameter}. For the parameters $X_2, X_4, X_5$ and $X_6$, three levels for the standard uncertainty are chosen: either 0, corresponding to no uncertainty, or 1, corresponding to the standard uncertainties given in Table~\ref{tab:parameter}, or 2, corresponding to the values of level 1 multiplied by a factor of 2. A benefit of setting a factor to three levels is that the main effect of the parameter is separated into the linear effect and the quadratic effect~\cite{connor1961fractional}, allowing one to observe if there are any significant non-linearities. For the parameter $X_3$, i.e., the deviation of the SUT from an assumed design, the three levels denote different choices of the actual deviation from the design. We set the levels to either 0, corresponding to no deviation, or 1, denoting the default deviation used throughout the previous sections and whose estimate can also be seen in Figure~\ref{fig:toroid} b), or 2, corresponding to the default deviation multiplied by a factor of 2, to simulate a larger deviation from the assumed design surface. Now, given the set of parameters and their possible levels, a design matrix is created that defines the individual runs. In our analysis, the experimental design we have opted for is a subset of the Taguchi L18 design~\cite{taguchi1987system}. This includes a design matrix, shown in Table~\ref{tab:designmat}, consisting of 18 runs of various level combinations for the 6 parameters $X_1$ to $X_6$.
A quality of the design chosen is that in every pair of columns, each possible level pairing occurs a constant number of times~\cite{Kacker1991toa}. That is, each level of a three-level parameter is paired with any level from a different three-level parameter the same number of times. Similarly, the higher and lower levels of the two level parameter ($X_1$) pair with each level of the three-level parameters on an equal number of occasions during the 18 runs. This negates the possibility of any interaction effects interfering with our analysis of the individual parameters. The reason behind this choice of design is hence that it covers the parameter space with (as with all fractional factorial designs) a lower variance of effects than a one-factor-at-a-time approach~\cite{czitrom1999one}, while also possessing the fewest number of runs where the orthogonality condition of constant level pairings holds, thus permitting valid conclusions to be drawn about the individual parameter effects in an economic manner. To avoid possible systematic effects, the order of the runs has been randomized.
\begin{figure}[htb]
    \centering
    \includegraphics[width=.99\linewidth]{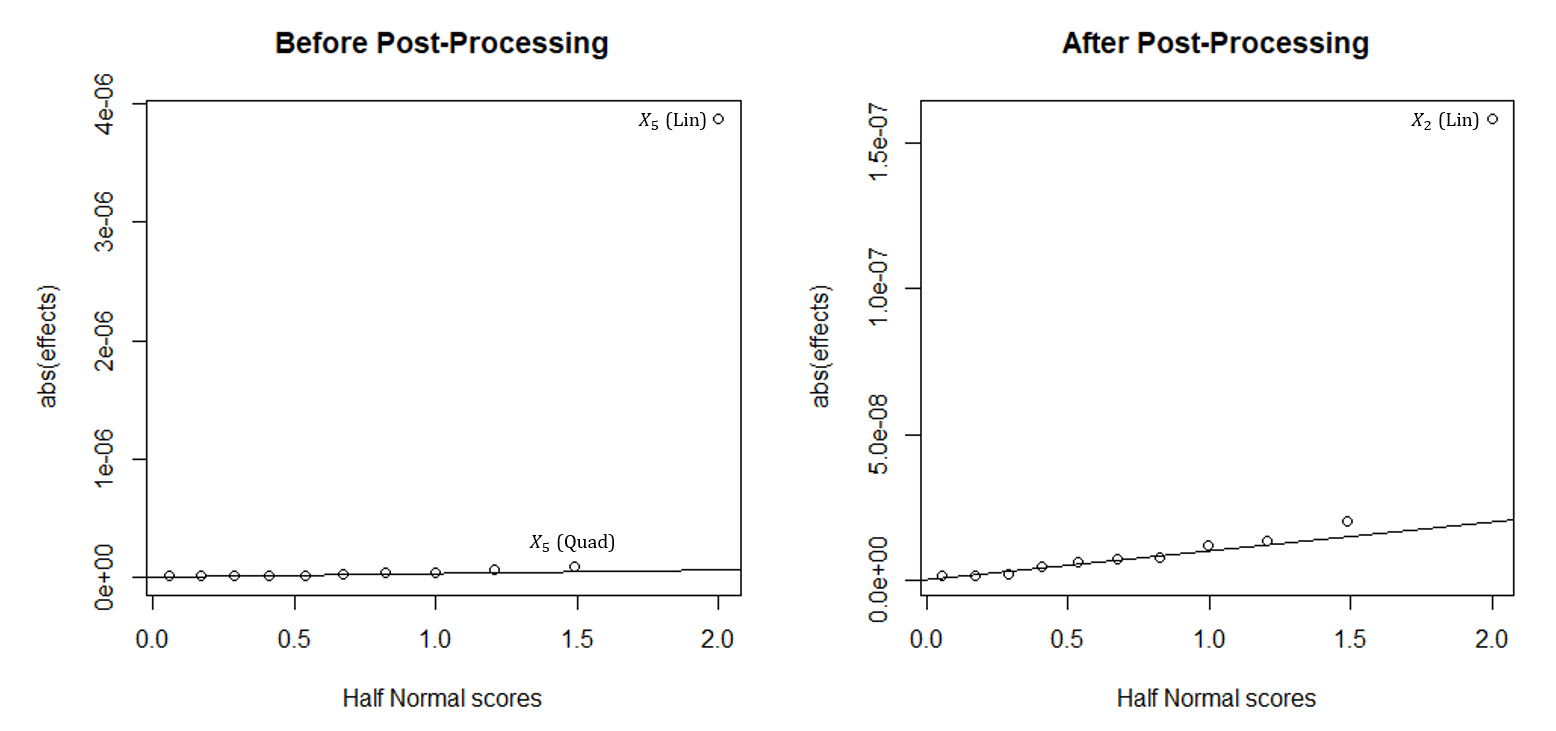}
    \caption{Half normal plots indicating the effect of the parameters on the uncertainty before and after the post-processing step. The parameters which are significantly influential at the 5\% level have been labelled, as well as whether they are linear (Lin) or quadratic (Quad) effects.}
    \label{fig:halfnormmat}
\end{figure}

The results of the experimental design, displayed in the half-normal plots in Figure~\ref{fig:halfnormmat}, highlight a sharp difference between the uncertainty evaluation before and after post-processing is applied. A half-normal plot illustrates the absolute effects of each parameter against their corresponding score from the half-normal distribution, and can identify influential effects by assessing whether the parameter deviates significantly (that is to say it cannot be justified as purely random deviation) from the straight regression line~\cite{scholz2022experimental}. In our analysis, the parameter $X_5$ ($(\Delta\alpha,\Delta\beta)$) was by far the main source of uncertainty before the post-processing step, while after post-processing, the parameter $X_2$ ($\Delta z$) was the predominant contributor identified by the design, with its linear effect the only term deemed significant. This is in keeping with our deductions regarding the influential factors in the TWI from section \ref{sec:numerics:asphere}, where the z-positioning was also found to be the main source of uncertainty after the orientation was corrected in the post-processing step. However, even in the TWI's linear model, it is worth noting that before post-processing, the quadratic effect of $X_5$ was also deemed significant at the 5\% level, in contrast to the linear effects of any other parameter. This suggests that it could be of use to consider and account for the uncertainty arising from non-linear terms in any future development of the TWI model. The main consideration for immediate work on the TWI, however, should center on reducing the uncertainty which arises from parameter $X_2$. Another result from the experimental design study is that larger deviations of the surface from the design, which are represented by parameter $X_3$, provide no significant contribution to the uncertainty - neither linear nor quadratic.
\begin{figure}[htb]
    \centering
    \includegraphics[width=.99\linewidth]{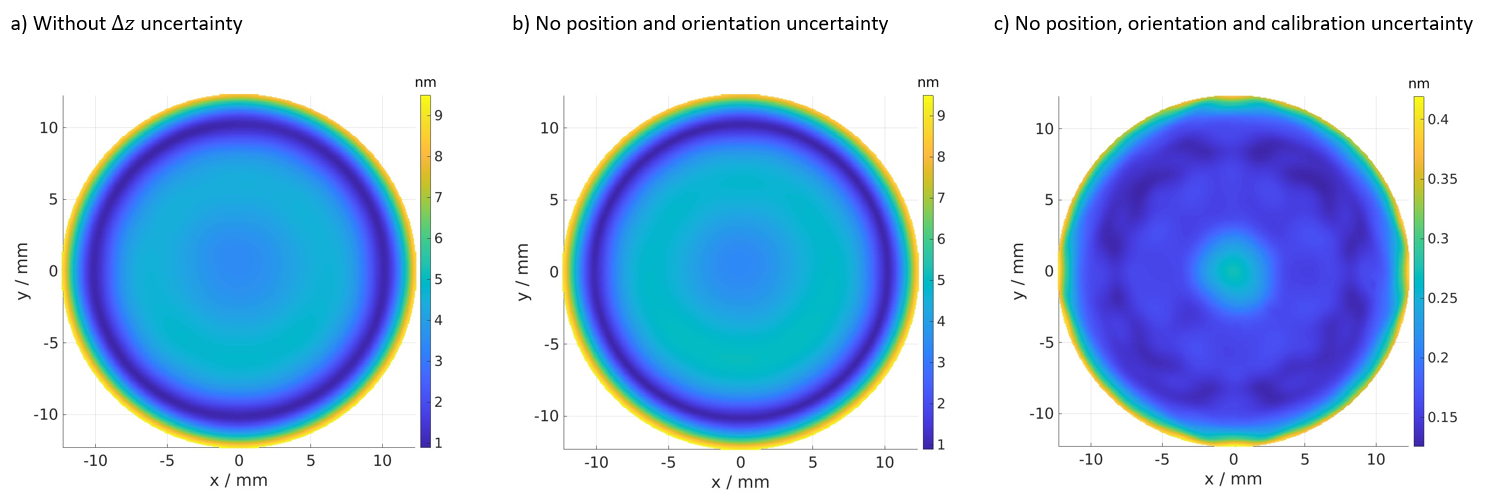}
    \caption{Standard uncertainty of the final surface form for the asphere is shown leaving out the uncertainty contribution of $\Delta z$ in a). In b), all position and orientation uncertainty contributions are ignored. In c) all position, orientation and calibration uncertainty contributions are ignored.}
    \label{fig:aspherewoliz}
\end{figure}

The statistical analysis above can also be constituted visually by repeating the uncertainty evaluation procedure while systematically leaving out certain parameters. For the asphere, in Figure~\ref{fig:aspherewoliz} a) the standard uncertainty of the absolute surface form is shown, when ignoring the uncertainty contribution of the main influencing factor $\Delta z$. The difference to Figure~\ref{fig:asphere} d) is clearly visible and the uncertainty caused by the spherical form deviation observed previously is removed, resulting in an rms value of 4.5 nm. This quantifies our conjecture and current work~\cite{Scholz_2023} aims for a significant reduction of the uncertainty in $\Delta z$. In Figure~\ref{fig:aspherewoliz} b) the pixel-wise standard uncertainty is shown for the case when assuming no uncertainties due to positioning and orientation. Comparing to the case before shows that the positioning and orientation (except for the positioning along the optical axis $\Delta z$) has almost no impact on the resulting final uncertainty. The reason is that the position and orientation of the SUT in a TWI (except for the z-positioning) can be deduced from the measurement data with high accuracy. We also show in Figure~\ref{fig:aspherewoliz} c) the resulting pixel-wise standard uncertainty after post-processing when applying the Bayesian uncertainty evaluation to the case of having no position, orientation and calibration uncertainty contribution. While the input uncertainty contributions are not considered in this case, the remaining term in the covariance approximation~\eqref{eq:covariance monte carlo}, namely $\sigma^2\left(J(\hat\theta, Z)^TJ(\hat\theta, Z)\right)^{-1}$, reflects remaining uncertainty due to the noise in the data and the model approximation. Quantitatively, this contribution admits a rms value of 0.2 nm. Also, the pattern in Figure~\ref{fig:asphere} c) shows an interesting structure which comes from the different patches used for surface reconstruction: Areas, where different patches overlap and where therefore the data point density is larger, show smaller uncertainty, which is a plausible result.

\begin{figure}[htb]
    \centering
    \includegraphics[width=.99\linewidth]{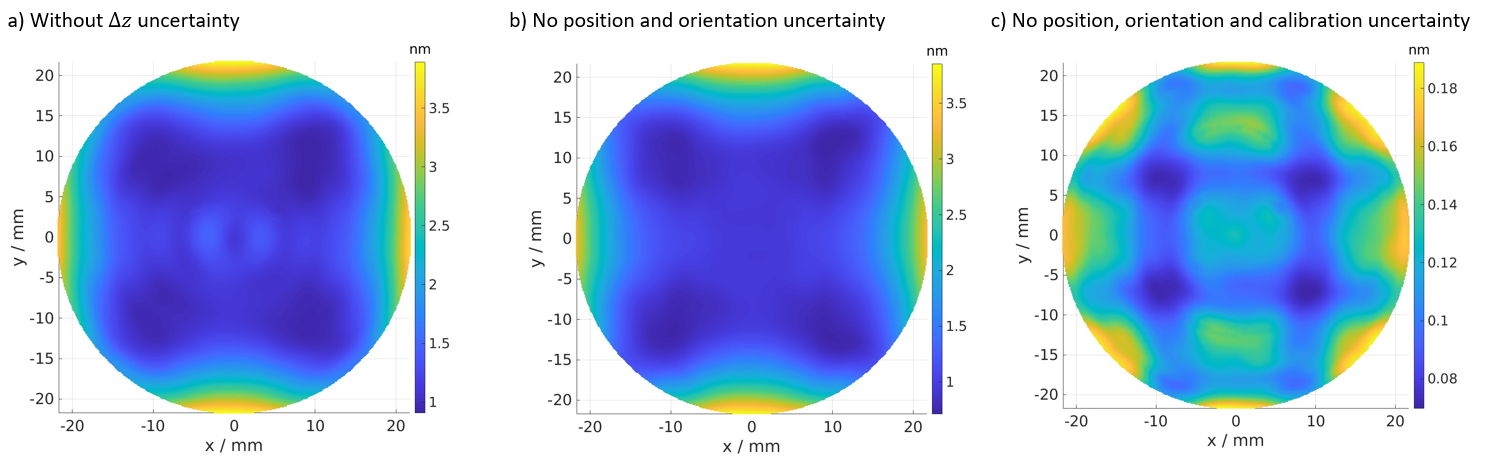}
    \caption{Standard uncertainty of the final surface form for the toroid is shown leaving out the uncertainty contribution of $\Delta z$ in a). In b), all position and orientation uncertainty contributions are ignored. In c) all position, orientation and calibration uncertainty contributions are ignored.}
    \label{fig:toroidwoliz}
\end{figure}

Similarly, Figure~\ref{fig:toroidwoliz} shows the same experiment for the toroid. Also here, the standard uncertainty of the absolute surface form is shown, when ignoring the uncertainty contribution of deviations in the $z$-direction. It can be seen by comparing Figure~\ref{fig:toroidwoliz} a) with Figure~\ref{fig:toroid} d), that the strong spherical effect is removed. This is even more visible in the rms values which, leaving out the effect of $\Delta z$, reduce to 2 nm for the final standard uncertainty of the absolute surface form. In Figure~\ref{fig:toroidwoliz} b) the standard uncertainty of the measurand is shown when neglecting all position and orientation uncertainty sources. Again, this has no significant effect, due to the measurement principles of the TWI. In Figure~\ref{fig:toroidwoliz}, additionally to orientation and positioning, the calibration uncertainty source is also ignored. This reduces the rms value to 0.1 nm and again, patch overlap areas are clearly visible and reduce the underlying uncertainty in areas where measurements are dense.

\section{Discussion and Outlook}
\label{sec:discussion}
In this work, a Bayesian uncertainty evaluation method has been developed, using an efficient approximation of the posterior PDF and the forward model. The efficacy of the approach was demonstrated on a case study based on the TWI, a relevant application from optical surface metrology, considering the computational model of the TWI and taking a number of key influencing factors into account. Reasonable and quantitative uncertainty estimates were obtained for the form of two typical surfaces (an asphere and a toroidal surface). An experimental design study confirmed the most important uncertainty sources and their contribution to the final standard uncertainty of the absolute surface form have been assessed. 

That is, to the best of our knowledge, the first work that proposes a Bayesian approach towards uncertainty estimation for the TWI. In particular, the approach presented efficiently deals with the complexity and high-dimensionality of the TWI and the embedded computational model that is required to mimic the optical properties of the TWI. However, the generality of the uncertainty evaluation approach presented renders it interesting for a variety of metrological applications. Moreover and in contrast to variance propagation approaches, the method presented allows for probability statements to be made, i.e., PDFs of single pixel, spatial correlations, and conformity properties can be analyzed.

Future research on the uncertainty evaluation approach presented may consider the incorporation of prior knowledge about the measurand, i.e. for the TWI, the parameterization of the surface form. This prior knowledge would allow the reduction of the resulting uncertainty by using additional expert knowledge and possibly historical measurements in the inference process. Moreover, the Bayesian approach can be used to validate the computational, together with the employed statistical model, and hence improve the computational model of the measurement process.
For the TWI application presented, future work may reduce the simplifications that were made to develop the methods. Furthermore, efforts are made to determine realistic values for the uncertainty contributions and investigate further impact factors and their influence on the uncertainty by applying the methods developed. Current research focuses on the determination and minimization of the most significant uncertainty source: the alignment of the surface under test along the optical axis of the interferometer. Finally, the results may be improved by investigating the process of the alignment of the surface under test within the interferometer setup and adding this procedure to the computational model.

%\section*{Acknowledgement}
\begin{backmatter}
\bmsection{Acknowledgements}
Financial support from the European Partnership on Metrology, co-financed from the European Union’s Horizon Europe Research and Innovation Programme and by the Participating States (Grant 22DIT01 ViDiT) is greatly acknowledged. The authors thank CC UPOB e.V. for providing the asphere measured in section~\ref{sec:numerics:asphere}.
% Funder name: European Partnership on Metrology 
% Funder ID: 10.13039/100019599 
% Grant number: 22DIT01 ViDiT
\bmsection{Disclosures}
The authors declare no conflicts of interest.
\bmsection{Data Availability Statement}

\end{backmatter}
\section*{Appendix}
\appendix

\section{Details on the parametrization of the computational model}
\label{sec:appendix:parameter}
In literature, cf. e.g.~\cite{fortmeier2022development}, the inference parameter $\theta$ of the least-squares regression problem~\eqref{eq:leastsquaresestimate} consists of a set of Zernike coefficients to represent design deviations of the SUT together with the design parameter. In particular, the corresponding Zernike polynomials describe a difference topography, which, together with an a priori chosen design topography, forms the SUT. Additionally, $\theta$ comprises the position and orientation of the SUT in the measurement system and a set of patch-offset values, which describe the unknown offsets of each patch of rays on the CCD, that are generated during the measurement. Depending on the actual form of the SUT, some of these parameters are highly or even perfectly correlated, for instance due to symmetry. This non-uniqueness and resulting non-identifiability of the non-linear regression problem had no effect on the quality of the estimation in previous works, since the employed Levenberg-Marquardt optimization procedure~\cite{doi:10.1137/0111030} regularizes the gradient updates. However, for the uncertainty evaluation scheme presented in section~\ref{sec:bayesian inference} and in algorithm~\ref{alg:mc}, the Jacobian of the least-squares functional is required to be of full rank. Therefore, the physical parameters corresponding to the orientation of the SUT, i.e., the $(\alpha, \beta, \gamma)$ rotation parameter, are removed. This renders the Jacobian of the least-squares functional full-rank. This adjustment to the inverse problem has no effect on the final result, since the positioning and orientation is optimized in the final post-processing step from section~\ref{sec:measurement}. For this subsequent fitting process, the projection to obtain higher-frequency terms of the SUT is not considered in this work, since the specimens considered do not admit higher-frequency terms due to their definition as smooth Zernike polynomials of the same degree that is used for reconstruction in~\eqref{eq:leastsquaresestimate}. Moreover, the higher-frequency projection introduces a dependence of the SUT parameter and the parameter of the optical system, e.g., the wavefront manipulators. This dependency is naturally treated using the Bayesian approach and future research is required to find a numerically efficient way to tackle this problem.

\section{Details on the calibration procedure and corresponding uncertainty evaluation}
\label{sec:appendix:calibration}

Similar to the measurement, i.e., the evaluation of the measurand, the calibration can be mathematically seen as an inverse problem. However, the role of the parameters in the CM are different. In particular, important properties of the traceable reference spheres are well-known and the Zernike coefficients of the wavefront manipulators are to be determined. 

For the CM defined in this work, the function $g(\theta, Z)$ takes as input the parameter $\theta$ mainly corresponding to the form of the SUT and the parameter $Z$ which reflects additional elements of the CM, required to replicate the TWI. Being more precise for the calibration task, the former can be decomposed in $\theta=(\theta_{\mathrm{SUT}}, \theta_{\mathrm{off}}, \theta_{\mathrm{pos}})$, where the individual (but still multivariate) parameters denote the parametrization of the form of the \emph{SUT}, the \emph{offset} values for each patch arising in the measurement and the \emph{positioning} parameter of the SUT in the instrument, respectively. The remaining parameter $Z$ can also roughly be decomposed in $Z=(Z_{\mathrm{wav}}, Z_{\mathrm{opt}})$, where $Z_{\mathrm{wav}}$ denotes the two sets of (double) Zernike coefficients~\cite{fortmeier2022development} for the \emph{wavefront manipulators} and $Z_{\mathrm{opt}}$ denotes the remaining properties of the \emph{optical} system (e.g., positioning, alignment of optical elements and their optical properties, the wavelength of the laser, etc.).

Due to the well-known properties of the reference spheres, $\theta_{\mathrm{SUT}}$ is assumed to be known. In fact, the values of $\theta_{\mathrm{SUT}}$ are given by the two radii of the reference spheres given in section~\ref{sec:numerics}. In practice, they are known with small uncertainties, which we assume to be negligible in this work. Then, given this parametrization and the real calibration measurements, the least-squares optimization problem for the calibration reads
\begin{equation}
    \label{eq:leastsquarescalibration}
    (\hat Z_{\mathrm{wav}}, \hat\theta_{\mathrm{off}}, \hat\theta_{\mathrm{pos}}) \in \argmin_{Z_{\mathrm{wav}}, \theta_{\mathrm{off}}, \theta_{\mathrm{pos}}} \|y_c - g((\theta_{\mathrm{SUT}}, \theta_{\mathrm{off}}, \theta_{\mathrm{pos}}), (Z_{\mathrm{wav}}, Z_{\mathrm{opt}}))\|^2_2,
\end{equation}
where $y_c$ denotes the OPDs acquired jointly for the reference spheres at all positions. Again, there is usually no analytical solution to~\eqref{eq:leastsquarescalibration} and numerical optimization is required. The estimation here is, however, even more complex and time consuming than the optimization of~\eqref{eq:leastsquaresestimate} to obtain the form of the SUT in a measurement. A reason for that is the huge amount of measurements contained in $y_c$, which corresponds to two reference spheres, measured at many different positions. For details on the calibration process, cf.~\cite{fortmeier2022development}. After the calibration has been performed, the calibration estimate $\hat Z_{\mathrm{wav}}$ can be employed for the estimation of an unknown SUT, as described in section~\ref{sec:measurement}.

For the uncertainty evaluation proposed in section~\ref{sec:bayesian inference}, a suitable prior distribution for the wavefront manipulator parameterization is required. This can be retrieved from the calibration using a Laplace approximation of the posterior in the estimate. For that, consider the calibration problem in a similar form as in~\eqref{eq:statistical model} using the abbreviation $\eta = (\hat Z_{\mathrm{wav}}, \hat\theta_{\mathrm{off}}, \hat\theta_{\mathrm{pos}})$ and $\zeta = (\theta_{\mathrm{SUT}}, Z_{\mathrm{opt}})$. Here, $\eta$ collects the unknown inference parameters and $\zeta$ the known (and fixed) parameters during calibration. With a slight abuse of notation, we write the CM now shorthand as $g_\zeta(\eta)$, where the parameters are plugged into the function $g$ according to~\eqref{eq:leastsquarescalibration}. Then, a suitable statistical model reads
\begin{equation}
    y_c \vert \eta, \zeta \sim N(g_\zeta(\eta), \sigma_c^2 I),\; \sigma_c^2>0 \quad\text{known}.
\end{equation}
The variance $\sigma_c^2$ during the calibration is again assumed to be known and we also consider the parameter $\zeta$ to be known and fixed. We assume, as above, the prior $\pi(\eta)$ to be non-informative, i.e., $\pi(\eta)\propto 1$. Then, the posterior according to Bayes' rule reads
\begin{equation}
  \label{eq:calib posterior}
    \pi(\eta \vert y_c, \zeta) \propto \exp\left(-\|g_\zeta(\eta) - y_c\|_2^2/2\sigma_c^2\right).
\end{equation}
The usual TWI estimation procedure during calibration yields approximately the MAP of this posterior, denoted here by $\hat\eta$. To additionally obtain information about the uncertainty, a linearization of $g_\zeta(\eta)$ can be performed around the estimate $\hat\eta$ 
\begin{equation}
    g_{\zeta}(\eta) \appropto g_\zeta(\hat\eta) + J_\zeta(\hat\eta)\left[\eta - \hat\eta\right].
\end{equation}
The Jacobian $J_\zeta$ evaluated at the estimate $\hat\eta$ can be derived analytically and evaluated numerically, cf.~\cite{fortmeier2014analytical}.
Inserting this linearization into the posterior yields a quadratic form in the exponent w.r.t. $\eta$. In particular it holds
\begin{align}
    -(2\sigma_c^2)^{-1}\|g_\zeta(\eta) - y_c \|_2^2 \approx& -(2\sigma_c^2)^{-1} \times \\ &\times\left\{\eta^T J_\zeta(\hat\eta)^T J_\zeta(\hat\eta) \eta -2\eta^TJ_\zeta(\hat\eta)^T\left(y_c - g_\zeta(\hat\eta) + J_\zeta(\hat\eta)\right) + \tilde{c}\right\}, \nonumber
\end{align}
where $\tilde{c}$ is a constant independent of $\eta$.Rearranging the terms in the exponent to fit the definition of a Gaussian distribution, $\Gamma:=\sigma_c^{2}(J_\zeta(\hat\eta)^TJ_\zeta(\hat\eta))^{-1}$ can be considered as the first order approximation to the covariance of the posterior~\eqref{eq:calib posterior}. Assuming a joint multivariate Gaussian distribution $\eta\sim N(\hat\eta, \Gamma)$, each multivariate collection of marginals is also a Gaussian distribution. Therefore, we extract the estimate $\hat Z_{\mathrm{wav}}$ and covariance $\Gamma_{\mathrm{wav}}$ from $\eta$ and $\Gamma$ by selecting the rows and columns that correspond to the wavefront manipulators. Note that the calibration data is taken from actual calibrations performed by the TWI at PTB Braunschweig.

\section{Derivation of the approximate covariance matrix for the measurement}
\label{sec:details covariance}
In section~\ref{sec:bayesian inference}, the posterior covariance~\eqref{eq:covariance} is approximated using a linearization of the CM in the measurand variable. This linearization is performed for every value of $Z$ drawn from the prior $\pi(Z)$ in a local optimal solution $\hat\theta:=\hat\theta(Z)$ of the problem~\eqref{eq:hat theta}. Hence, for $\pi(Z)$-almost every $Z$, the joint posterior~\eqref{eq:joint posterior} can be approximated by a normal distribution in the variable $\theta$. In particular, it holds
\begin{equation}
  \label{eq:Laplace approximation}
    \pi(\theta, Z \; \vert\; y) \approx \phi(\theta; \hat\theta, \Gamma)\pi(Z),
\end{equation}
where $\phi$ denotes the PDF of a normal distribution with mean $\hat\theta$ and covariance $$\Gamma=\sigma^{2}(J(\hat\theta, Z)^TJ(\hat\theta, Z))^{-1}$$. Note that $\Gamma$ in this section corresponds to the covariance used in the measurement stage, which is different from the covariance used in the calibration stage of appendix section~\ref{sec:appendix:calibration}. The actual form for the covariance is an immediate consequence of the linearization of the CM and a Laplace approximation that preserves the local curvature of the posterior PDF at the expansion point. See also~\ref{sec:appendix:calibration} for a details on this argument. With this, the integration over the variable $\theta$ in the formula for the covariance~\eqref{eq:covariance} can be performed analytically. In particular, it holds
\begin{equation*}
    U \approx \int (\theta-\hat\theta_{\hat Z})(\theta-\hat\theta_{\hat Z})^T\int \phi(\theta, \hat\theta,\Gamma)\;\mathrm{d}\pi(Z)\;\mathrm{d}\pi(\theta).
\end{equation*}
Using Fubini's Theorem, one can interchange the integration order and the constant prior $\pi(\theta)\propto 1$ yields
\begin{align}
  \label{eq:covariance terms}
    U&\approx \int \int \theta\theta^T \phi(\theta; \hat\theta, \Gamma)\mathrm{d}\theta + \int \hat\theta_{\hat Z}\hat\theta_{\hat Z}^T \phi(\theta; \hat\theta, \Gamma)\mathrm{d}\theta \\
    &\qquad - \hat\theta_{\hat Z}\int \theta^T\phi(\theta; \hat\theta, \Gamma)\;\mathrm{d}\theta\;\mathrm{d}\pi(Z) - \int \theta\phi(\theta; \hat\theta, \Gamma)\;\mathrm{d}\theta\;\mathrm{d}\pi(Z)\hat\theta_{\hat Z}^T.
\end{align}
The first term in~\eqref{eq:covariance terms} is the second moment of the multivariate normal distribution $N(\theta; \hat\theta, C)$, i.e., it equals its covariance plus the squared first moment, $\Gamma + \hat\theta\hat\theta^T$. For the second term in~\eqref{eq:covariance terms}, it is to note that $\hat\theta_{\hat Z}$ does not depend on $\theta$, hence the second term is an integral over a PDF, which equals one, times $\hat\theta_{\hat Z}\hat\theta_{\hat Z}^T$. And the integrals in the third and fourth term are simply the mean of the normal distribution $N(\theta; \hat\theta, \Gamma)$. Therefore, it holds
\begin{equation}
    U \approx \int \Gamma + \hat\theta\hat\theta^T + \hat\theta_{\hat Z}\hat\theta_{\hat Z}^T - \hat\theta_{\hat Z}\hat\theta^T - \hat\theta\hat\theta_{\hat Z}^T\;\mathrm{d}\pi(Z) = \int \Gamma + (\hat\theta_{\hat Z} - \hat\theta)(\hat\theta_{\hat Z} - \hat\theta)^T \;\mathrm{d}\pi(Z).
\end{equation}
Since $\Gamma = \sigma^2 (J(\hat\theta, Z)^TJ(\hat\theta, Z))^{-1}$, the formula~\eqref{eq:covariance monte carlo} is derived, where the quality of the approximation depends on the quality of the approximation to the joint posterior PDF~\eqref{eq:Laplace approximation}.

\section{Design Matrix}
\label{sec:design matrix}
The design matrix for the experimental design in section~\ref{sec:numerics:sensitivity} consists of 18 runs for the 6 parameters $X_1,...,X_6$. Table~\ref{tab:designmat} displays the level each parameter was set to for the respective run number. The lowest level for each parameter corresponds to 0, with the next highest being represented by a 1. In the case of the three-leve parameters ($X_2,...,X_6$), the maximum level is selected where a 2 is listed. One can observe how each possible level pairing occurs an equivalent number of times, which ensures interactive effects will not influence the individual parameter effects.
\begin{table}[htb]
\centering
\caption{The design matrix for the experimental design in section~\ref{sec:numerics:sensitivity}.}
\begin{tabular}{|c|c c c c c c|}
  \hline
 & $X_1$ & $X_2$ & $X_3$ & $X_4$ & $X_5$ & $X_6$ \\ 
  \hline
1 & 0 & 2 & 2 & 2 & 2 & 2 \\ 
  2 & 0 & 1 & 1 & 2 & 2 & 0 \\ 
  3 & 1 & 0 & 1 & 2 & 0 & 1 \\ 
  4 & 0 & 2 & 0 & 2 & 1 & 1 \\ 
  5 & 0 & 0 & 0 & 1 & 1 & 2 \\ 
  6 & 1 & 2 & 1 & 1 & 0 & 2 \\ 
  7 & 0 & 1 & 2 & 1 & 0 & 0 \\ 
  8 & 0 & 0 & 0 & 0 & 0 & 0 \\ 
  9 & 1 & 0 & 2 & 1 & 2 & 1 \\ 
  10 & 1 & 2 & 1 & 0 & 1 & 0 \\ 
  11 & 0 & 2 & 2 & 0 & 0 & 1 \\ 
  12 & 1 & 0 & 2 & 2 & 1 & 0 \\ 
  13 & 1 & 1 & 0 & 2 & 0 & 2 \\ 
  14 & 1 & 2 & 0 & 1 & 2 & 0 \\ 
  15 & 1 & 1 & 0 & 0 & 2 & 1 \\ 
  16 & 0 & 1 & 1 & 1 & 1 & 1 \\ 
  17 & 1 & 1 & 2 & 0 & 1 & 2 \\ 
  18 & 0 & 0 & 1 & 0 & 2 & 2 \\ 
   \hline
\end{tabular}
\label{tab:designmat}
\end{table}

\bibliography{references}
\end{document}